%% file: apssamp.tex
\documentclass[%
 reprint,
 amsmath,amssymb,
 aps,
]{revtex4-2}

\usepackage{graphicx}
\usepackage{dcolumn}
\usepackage{bm}
\usepackage{multirow}%
\usepackage{amsmath,amssymb,amsfonts}%
\usepackage{amsthm}%
\usepackage{physics}
\usepackage{float}
\usepackage{lineno}
\usepackage{color}

\usepackage[hidelinks]{hyperref}

\begin{document}

\preprint{APS/123-QED}

\title[Increased light-emission efficiency in disordered InGaN through the correlated reduction of recombination rates]{Increased light-emission efficiency in disordered InGaN through the correlated reduction of recombination rates}

\author{Nick Pant}
\email{nickpant@umich.edu}
\affiliation{Applied Physics Program, University of Michigan, Ann Arbor, MI, USA 48109}
\affiliation{Department of Materials Science \& Engineering, University of Michigan, Ann Arbor, MI, USA 48109}
\author{Emmanouil Kioupakis}%
\affiliation{Department of Materials Science \& Engineering, University of Michigan, Ann Arbor, MI, USA 48109}

\date{\today}

\begin{abstract}
\noindent Experiments have shown that the light-emission efficiency of indium gallium nitride (InGaN) light-emitting diodes improves with increasing indium concentration. It is widely thought that compositional fluctuations due to indium incorporation suppress diffusion of carriers to non-radiative centers, thus leading to defect-insensitive emission. However, recent experiments have challenged this hypothesis by revealing unexpectedly long diffusion lengths at room temperature. Here, we demonstrate an alternative mechanism involving the correlated reduction in radiative and non-radiative recombination rates that explains the increase in light-emission efficiency of InGaN with increasing indium concentration, without invoking the suppression of carrier diffusion. Our analysis challenges the notion that carrier localization gives rise to defect tolerance in InGaN. 
\end{abstract}

\maketitle

\input{main}

\bibliography{Reference}

\end{document}

%% file: main.tex
\section{Introduction}

Shockley-Read-Hall (SRH) recombination is a fundamental loss mechanism that impairs the energy efficiency of all light-emitting materials \cite{shockleyStatisticsRecombinationsHoles1952, hallElectronHoleRecombinationGermanium1952}. In this process, electrons and holes recombine non-radiatively over crystalline defects to produce heat rather than recombining radiatively to produce light. The tolerance of a semiconductor to defects plays a key role in determining its suitability for light-emitting diodes (LEDs). Notably, indium gallium nitride (InGaN) LEDs exhibit remarkable brightness, even in the presence of defect densities surpassing $10^{10}$ cm$^{-2}$, six orders of magnitude higher than those found in III-V arsenide and phosphide emitters \cite{nakamuraCandelaClassHigh1994, lesterHighDislocationDensities1995, mukaiInGaNBasedBlueLightEmitting1998, chichibuLimitingFactorsRoomtemperature2005}. This is particularly advantageous considering InGaN serves as the most common active region in commercial visible LEDs \cite{pimputkarProspectsLEDLighting2009}.

Despite their resounding success, a comprehensive understanding of the performance of nitride LEDs remains incomplete \cite{weisbuchDisorderEffectsNitride2021}, limiting their advancement into longer visible and shorter ultraviolet wavelengths \cite{wasisto_beyond_2019, kneisslEmergenceProspectsDeepultraviolet2019} that are crucial for augmented-reality and germicidal applications. Experiments have shown that increasing the compositional disorder of InGaN increases the internal quantum efficiency (IQE), which is the ratio of the radiative recombination rate to the total recombination rate \cite{chichibuSpontaneousEmissionLocalized1996, narukawaRoleSelfformedInGaN1997, chichibu2001impact, onuma2005localized, chichibuOriginDefectinsensitiveEmission2006}, as reproduced in Figure 1. (At higher indium concentrations, the IQE drops because of a degradation in the material quality.) The improvement in efficiency with indium concentration is widely used to justify the claim that alloy disorder in InGaN imbues it with some tolerance to defects. However, care must be taken in interpreting these IQE measurements, as they were obtained using techniques that do not give access to the operating carrier density in the active region. Here, we propose an alternative mechanism relating to the operating carrier density that can explain the experimentally observed increase in the IQE without the notion of defect tolerance.

\begin{figure}[htp]
    \centering
    \includegraphics[width=0.4\textwidth]{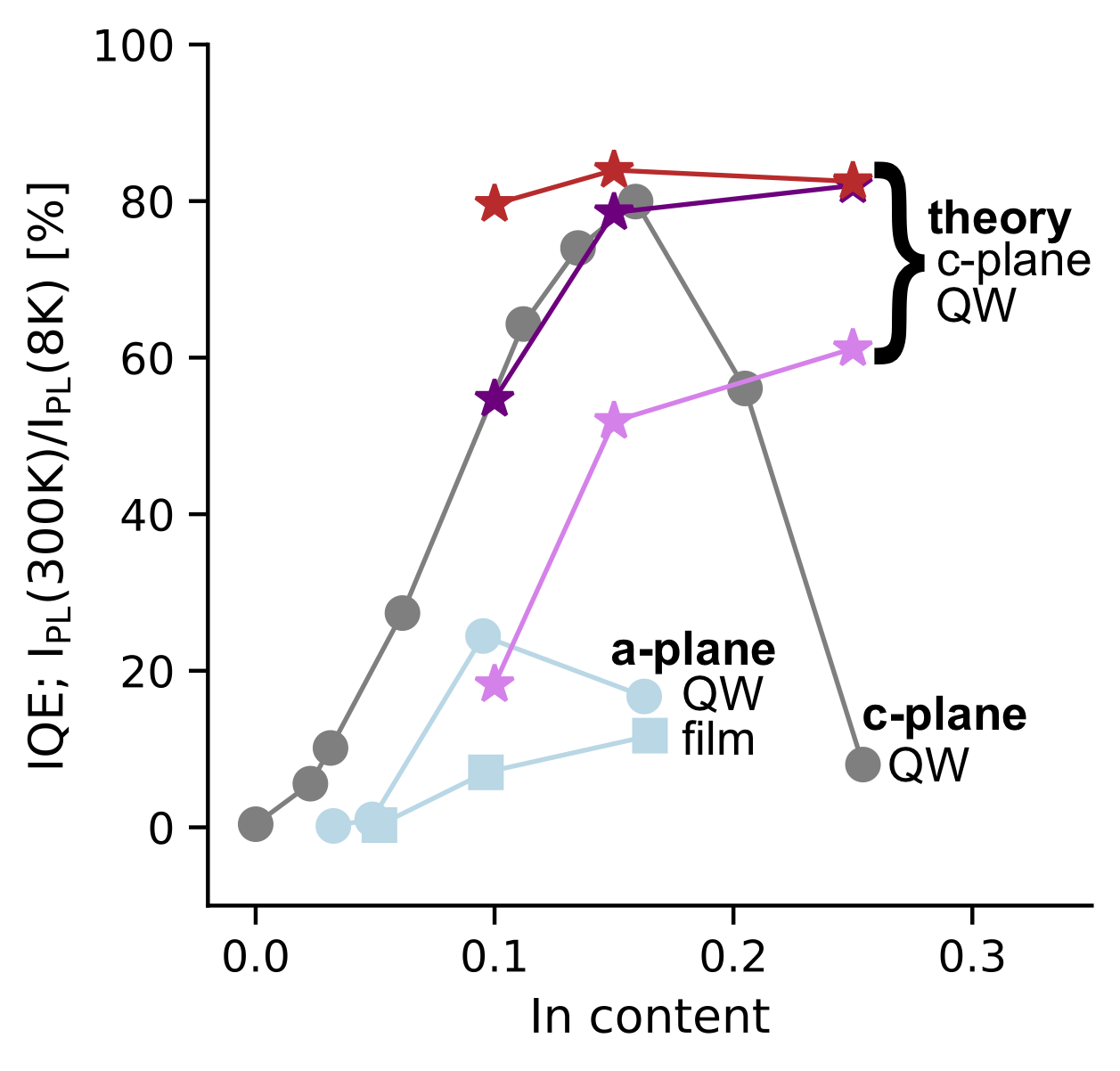}
    \caption{An increase in the IQE with increasing indium concentration is often used to justify the notion that carrier localization leads to defect tolerance in InGaN by suppressing diffusion to non-radiative centers (see, \textit{e.g.}, Ref. \cite{chichibuOriginDefectinsensitiveEmission2006}). The experimental points (blue circles and squares, and gray circles) were obtained from references \cite{chichibu2001impact, onuma2005localized, chichibuOriginDefectinsensitiveEmission2006}. We propose an alternate explanation that also predicts an increase in the IQE with increasing indium concentration without invoking the suppression of carrier diffusion to non-radiative centers (denoted ``theory"). We calculated the pink, purple, and red data points for recombination current densities of 1, 10, and 100 mA/cm$^2$.}
    \label{fig:def-tol}
\end{figure}

According to the prevailing hypothesis, compositional fluctuations in InGaN confines carriers to randomly occurring regions of narrower band gap and higher crystal quality 
\cite{chichibuSpatiallyResolvedCathodoluminescence1997, odonnellOriginLuminescenceInGaN1999, narukawaRoleSelfformedInGaN1997, bellaicheResonantHoleLocalization1999, oliverMicrostructuralOriginsLocalization2010}, preventing them from diffusing towards non-radiative recombination centers \cite{chichibuSpontaneousEmissionLocalized1996, chichibuOriginDefectinsensitiveEmission2006, chichibuReviewDefectTolerantLuminescent2019}. Following this reasoning, one might expect the internal quantum efficiency (IQE) of InGaN LEDs, defined as the ratio of the radiative recombination rate to the total recombination rate, to be relatively independent of defect density; however, empirical evidence contradicts this notion \cite{schubertEffectDislocationDensity2007, monemarDefectRelatedIssues2007, haderDensityactivatedDefectRecombination2010, haderTemperaturedependenceInternalEfficiency2011, armstrongDependenceRadiativeEfficiency2012, armstrongDefectreductionMechanismImproving2015,hallerBuryingNonradiativeDefects2017, hallerGaNSurfaceSource2018, polyakovEffectsMeVElectron2020, pivaDefectIncorporationIncontaining2020, roccatoEffectsQuantumwellIndium2021}. Even the hypothesis that disorder suppresses diffusion in InGaN has come under scrutiny, because a substantial fraction of carriers at room temperature are extended and able to diffuse \cite{davidManyBodyEffectsStrongly2019}. Diffusion lengths in InGaN alloys reach tens of microns at room temperature \cite{davidLongRangeCarrierDiffusion2021, shenThreeDimensionalModelingMinorityCarrier2021, nomeikaImpactCarrierDiffusion2022, bechtDiffusionAnalysisCharge}, surpassing the typical distance between defects. This poses perplexing questions: does InGaN exhibit defect tolerance and what role does disorder play?

The influence of compositional disorder on recombination in InGaN remains a subject of intense debate
\cite{kawakamiOriginHighOscillator2006, yangInvestigationNonthermalMechanism2008, humphreysDoesFormInrich2007, langerOriginGreenGap2011, karpovEffectCarrierLocalization2018, hammersleyConsequencesHighInjected2012, crutchleyInfluenceTemperatureRecombination2013, badcockCarrierDensityDependent2013, balochRevisitingInclusteringQuestion2013, choEfficiencyDroopLightemitting2013, yangInfluenceRandomIndium2014, schulzAtomisticAnalysisImpact2015, jeongCarrierLocalizationInrich2015, aufdermaurEfficiencyDropGreen2016, tannerAtomisticAnalysisElectronic2016, jonesImpactCarrierLocalization2017, liLocalizationLandscapeTheory2017, shahmohammadiEnhancementAugerRecombination2017, hahnEvidenceNanoscaleAnderson2018, christianRecombinationPolarInGaN2018, blenkhornResonantPhotoluminescenceStudies2018, karpovEffectCarrierLocalization2018, davidReviewPhysicsRecombinations2019, tannerPolarGaGa2020, kazazisTuningCarrierLocalization2020, davidExcitonsDisorderedMedium2022, sautyLocalizationEffectPhotoelectron2022}. For example, the tolerance of InGaN quantum wells to threading dislocations has been explained in terms of energetic shielding of dislocation cores due to quantum confinement. While this hypothesis certainly explains why high threading dislocation densities are not catastrophically detrimental to  InGaN LEDs, it fails to account for the increased efficiency of disordered bulk films with increasing composition (\textit{e.g.}, see \textit{a}-plane film in Figure 1) \cite{hangleiterSuppressionNonradiativeRecombination2005, abellRoleDislocationsNonradiative2008, quanRolesVshapedPits2014}. Another hypothesis posits that phase-segregation of cations \cite{massabuauCarrierLocalizationVicinity2017} prevents carriers from reaching dislocation cores, but this explanation does not account for non-radiative recombination involving point defects away from dislocations, which contribute significantly to non-radiative recombination \cite{alkauskas2016role}. Studying the impact of compositional disorder on recombination dynamics is challenging experimentally. The difficulty lies in independently controlling disorder while keeping other influential factors, such as growth conditions, microscopic polarization fields, and defect-transition levels, constant. On the other hand, it is much easier to control for disorder effects with predictive numerical-modeling techniques. By explicitly solving the Schr\"{o}dinger equation, which governs the physics of electrons and holes, we can disentangle the impact of localization on recombination from other factors. 

In this work, we investigate the claim that carrier localization is responsible for the defect-insensitive emission of InGaN. To test this hypothesis, we simply calculate what happens to the competition between radiative and SRH recombination if we change the degree of localization. By manipulating the \textcolor{black}{alloy-scattering potential} of the random alloy, we directly control the strength of alloy disorder and solve the Schr\"{o}dinger and Poisson equations in the modified potentials. \textcolor{black}{Our analysis includes hundreds of eigenstates, spanning the range from localized to extended.} Our results challenge the status quo, and we find that carrier localization does \textit{not} increase the ratio $B/A$ or the IQE for a given defect density. Instead, we propose that carrier localization slows both radiative and non-radiative recombination in a correlated manner; this increases the operating carrier density and promotes bimolecular radiative recombination over monomolecular non-radiative recombination for a given operating current density or optical illumination intensity. \textcolor{black}{Therefore, we propose that the enhancement of the IQE in Figure 1 is very likely be due to differences in the carrier density with increasing carrier localization rather than due to defect tolerance.} Additionally, because polarization fields in quantum wells also increase the operating carrier density, they have a similar effect in promoting radiative recombination over non-radiative recombination in \textit{c}-plane quantum wells at low currents. 

\section{Theoretical modeling of recombination}

\subsection{Schr\"{o}dinger-Poisson calculations}

To assess the influence of compositional disorder on the competition between radiative and SRH recombination, we have developed a theoretical framework based on solutions to the Schr\"{o}dinger and Poisson equations. In our approach, we model alloy disorder in InGaN by randomly assigning the composition $x$ in a grid as either $x=0$ (GaN) or $x=1$ (InN). By manipulating the \textcolor{black}{alloy-scattering potential}, we directly control the degree of carrier localization, which allows us to probe its impact on recombination dynamics. For our calculations, we used first-principles density-functional theory to parameterize the effective-mass $k \cdot p$ Hamiltonian \cite{rinkeConsistentSetBand2008, dreyerEffectsStrainElectron2013, wrightElasticPropertiesZincblende1997, yanEffectsStrainBand2014, dreyerCorrectImplementationPolarization2016, mosesHybridFunctionalInvestigations2011}, which we diagonalized using \texttt{nextnano++} \cite{birnerNextnanoGeneralPurpose2007}, \textcolor{black}{obtaining hundreds of eigenstates}. We modeled periodic supercells of size 18 nm $\times$ 18 nm $\times$ 18 nm with a grid-size spacing (0.3 nm) that corresponds to the nearest-neighbor distance between cations in InGaN. Unless specified otherwise, we repeated each calculation for ten different configurations of the random alloy. \textcolor{black}{We tuned the hole scattering potential $\Delta E_v$ to span the range from 0.0 eV to 1.0 eV, while fixing the electron scattering potential $\Delta E_c$ to the first-principles conduction-band offset energy of 2.3 eV} \cite{mosesHybridFunctionalInvestigations2011}. We note that experimental measurements have determined the VB offset to be between 0.5 eV and 1.1 eV \cite{martinValenceBandDiscontinuities1996, shihBandOffsetsInN2005, ohashiHighStructuralQuality2006, wuCrosssectionalScanningPhotoelectron2008, mahmoodDeterminationInNGaN2007, wuPolarizationinducedValencebandAlignments2007, wangConductionBandOffset2007}. \textcolor{black}{To better understand the meaning of the scattering potentials, it is illustrative to consider the extreme limit of a virtual-crystal alloy, for which there is no alloy-scattering potential, and $\Delta E_v = \Delta E_c = 0$.}

\subsection{Relation between recombination rates and wave-function overlaps}

Central to our analysis is the relationship between recombination rates and the overlap of the carrier wave functions from $k \cdot p$ perturbation theory. The rate of radiative recombination is proportional to the squared overlap of the electron and hole wave functions $\abs{F_{eh}}^2$, \cite{foxOpticalPropertiesSolids2001, jonesImpactCarrierLocalization2017, davidCompensationRadiativeAuger2019, liCarrierDynamicsBlue2023} 
\begin{linenomath}
\begin{align}
    \abs{\tilde F_{eh}}^2 &= \frac{\sum_{c,v} f_c (1-f_v) \abs{\int d^3\textbf{r} \psi_c(\textbf{r}) \psi_v(\textbf{r}) }^2}{\sum_{c,v}f_c(1-f_v)} \\
    \abs{F_{eh}}^2 &= \abs{\tilde F_{eh}}^2/\abs{\tilde F^{\text{VCA}}_{eh}}^2, \label{eq:Feh2}
\end{align}
\end{linenomath}
where $\psi$ is the envelope wave function, $f$ is the non-degenerate occupation probability at room temperature, and the indices $c$ and $v$ correspond to conduction band (CB) and valence band (VB) states, respectively. \textcolor{black}{We emphasize that the summations over $c$ and $v$ are performed over hundreds of eigenstates, including extended hole states, and we ensure that including more eigenstates in our solver does not appreciably change the overlap.} Equation (\ref{eq:Feh2}) rescales the overlap with respect to the virtual-crystal approximation (VCA) to ensure the limit $\abs{F_{eh}}^2=1$ for systems with translational invariance. \textcolor{black}{Therefore, $\abs{F_{eh}}^2 > 1$ being greater than one means that the overall probability of spontaneous recombination is exacerbated compared to a virtual-crystal, while $\abs{F_{eh}}^2 < 1$ means that the probability is suppressed.} $\abs{F_{eh}}^2$ is proportional to the radiative recombination coefficient $B$, which relates to the rate of radiative recombination $R_{\text{rad}}$ according to $R_{\text{rad}}=Bn^2$, where $n$ is the carrier density. The quantity $\abs{F_{eh}}^2$ is proportional to the oscillator strength, which quantifies the probability of light emission from a material.

The rate of SRH recombination is also proportional to an integral that resembles an overlap, 
\begin{linenomath}
\begin{equation}
    \abs{F_{SRH}}^2 = V(1+\kappa)\int d^3\textbf{r} \frac{\delta n(\textbf{r}) \delta p(\textbf{r})}{\delta n(\textbf{r}) + \kappa \delta p(\textbf{r})}, \label{eq:FSRH2}
\end{equation}
\end{linenomath}
where $\kappa \equiv c_p/c_n$ is the ratio of a given defect’s hole and electron capture coefficients, and $\delta n(\textbf{r}) \equiv n(\textbf{r})/N$ and $\delta p(\textbf{r}) \equiv p(\textbf{r})/N$ are the electron and hole densities divided by the total number of carriers $N$ in the simulation volume $V$. We provide a derivation of equation (\ref{eq:FSRH2}) in Appendix \ref{app:derivation}, \textcolor{black}{and discuss how the microscopic details of the atomistic physics involving carrier capture by defects, multi-phonon-emission, and details of the defect wave functions, are embedded in the $\kappa$ parameter through the capture coefficients.} The term $\abs{F_{SRH}}^2$ is proportional to the SRH recombination coefficient $A$, which in turn is related to the SRH rate according to $R_{\text{SRH}} = An$. One can also check that $\abs{F_{SRH}}^2=1$ for systems with translational invariance. \textcolor{black}{We note that $\abs{F_{SRH}}^2$ does not depend on $c_n$ and $c_p$ independently, and only depends on their ratio $\kappa$, representing a convenient separation of physics by length scale.} Very large or very small values of $\kappa$ indicate poor non-radiative recombination limited by the capture of either electrons or holes, respectively, meaning that these two processes are decoupled. In contrast, values of $\kappa$ near unity indicate that a defect captures electrons and holes successively and quickly, thus the probability of recombination depends strongly on the probability of both carriers being found at the defect site at the same time. Thus, by studying how compositional disorder impacts $\abs{F_{eh}}^2$ and $\abs{F_{SRH}}^2$, we can evaluate its influence on the radiative and SRH recombination rates.

\textcolor{black}{\subsection{Connection to approaches that explicitly model diffusion}}

\textcolor{black}{Since carriers need to diffuse to defects in order for them to be captured by multi-phonon emission and diffuse to each other in order for them to spontaneously recombine, one may wonder if it is necessary to explicitly model the dynamic evolution of carriers to accurately model the quantum efficiency. Fortunately, information of the time evolution of a system (in the absence of an external driving force) is embedded in its energy eigenstates \cite{sakurai1995modern}, and the challenge is in simply extracting this information. For a system in quasi-steady-state equilibrium, the problem is simplified because the system is accurately described as a linear combination of its energy eigenstates, weighted by thermal occupation factors (in this case, determined using Fermi-Dirac statistics) \cite{ridley2013quantum}. In this limit, the rate of dynamic processes can be efficiently computed using Fermi's golden rule, which involves integration of wave-function overlaps over the eigenspectra of the Hamiltonian \cite{sakurai1995modern, ridley2013quantum}. With respect to this work, it is the wave-function overlaps that contain information about whether a carrier can diffusively reach a defect. For example, a hole wave function localized away from a defect will have negligible overlap with the defect state, while a delocalized wave function that can reach the defect will have finite overlap with it. The formalism based on energy eigenstates has been successfully and widely applied to study both radiative and non-radiative processes in nitride semiconductors, using density-functional theory \cite{alkauskasFirstprinciplesTheoryNonradiative2014, kioupakisFirstprinciplesCalculationsIndirect2015}, empirical tight-binding \cite{mcmahonAtomisticAnalysisRadiative2020, mcmahonAtomisticAnalysisAuger2022}, and Schr\"{o}dinger-Poisson solvers \cite{kioupakisInterplayPolarizationFields2012, yangInfluenceRandomIndium2014, jonesImpactCarrierLocalization2017, davidFieldassistedShockleyReadHallRecombinations2017, davidCompensationRadiativeAuger2019, davidManyBodyEffectsStrongly2019}.}

\textcolor{black}{We briefly discuss the challenges present in other approaches that explicitly model the dynamic evolution of carriers, and justify our choice of employing the eigenstate formalism. The most general approach involves non-adiabatically propagating carrier populations and coherences in time, while simultaneously evolving the lattice \cite{curchod2018ab}. The advantage of this method is that it is valid out of equilibrium and can capture non-adiabatic processes such as hopping of localized states. However, this approach is intractable for the problem at hand as it is limited to small supercells that do not capture carrier localization, and the required timescales would span order of magnitudes from femtoseconds for the electron-phonon interaction to nanoseconds and microseconds for recombination. Another approach couples the Schrödinger and Poisson equations (or the Localization Landscape model) with the drift-diffusion equation \cite{o2021atomistic, shen2021three, li2017localization}. The drift-diffusion equation is derived from Boltzmann transport theory, which by design only tracks carrier populations and neglects wave coherences \cite{ponce2020first}. Applying the drift-diffusion model to the study of localized states is not justified from a theoretical standpoint because localization is fundamentally a manifestation of wave coherences due to multiple scattering \cite{anderson1958absence}. One may, of course, correct the diffusion constant and mobility to account for the fact that a fraction of carriers are localized, but this is only valid if one is interested in studying \textit{average} transport effects rather than the behaviour of localized carriers. From a technical perspective, most Schr\"{o}dinger-Poisson-drift-diffusion models also suffer from the self-interaction problem arising from the use of the Hartree approximation \cite{martin2020electronic}, which artificially delocalizes wave functions, which we have observed in drift-diffusion simulations that we performed using \texttt{nextnano++}. In contrast, the approach based on energy eigenstates that we have adopted in this work does not suffer from these technical issues, and provides a way to accurately model the quantum efficiency under quasi-steady-steady conditions.}

\section{Results and discussion} 

\subsection{Localization of wave functions}

\begin{figure}[htp]
    \centering
    \includegraphics[width=0.5\textwidth]{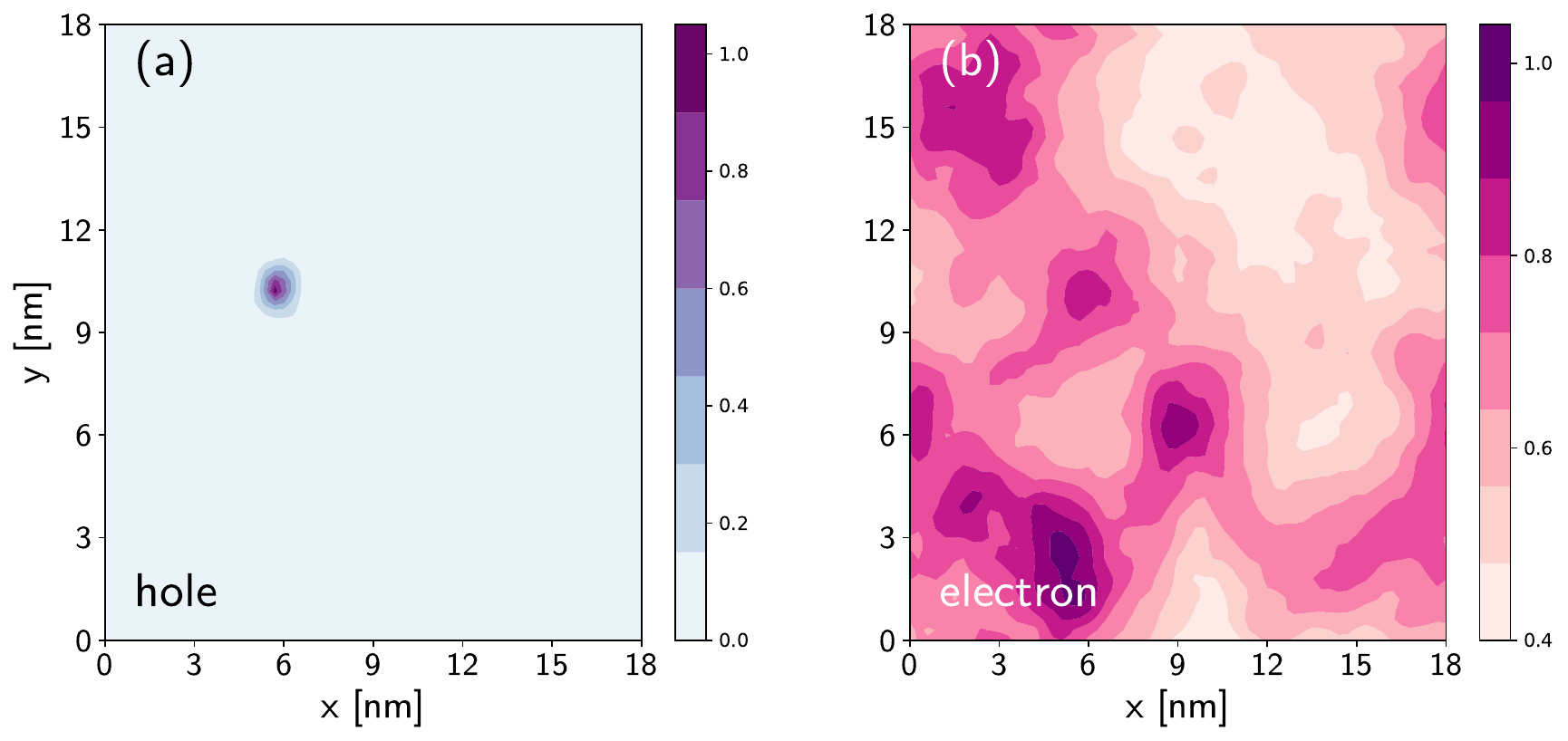}
    \caption{Carrier localization by compositional disorder in InGaN. Squared modulus of the ground-state (a) hole and (b) electron envelope wave functions, averaged along the \textit{c} direction, of an In$_{0.15}$Ga$_{0.85}$N alloy calculated with first-principles material parameters. Holes are strongly localized with a characteristic length scale of $\sim$1 nm while electrons are extended. The wave functions are rescaled so that their peak value is one.}
    \label{fig:1}
\end{figure}

Our solutions to the Schr\"{o}dinger and Poisson equations indicate that compositional disorder in InGaN leads to the localization of carrier wave functions. This localization stems from the confinement of wave functions in regions of narrower band gap as well as the interference of wave functions due to multiple scattering by the disordered potential \cite{filocheLocalizationLandscapeTheory2017}. We find that while electrons exhibit extended wave functions, holes near the valence-band maximum are strongly localized (Figure \ref{fig:1}), consistent with atomistic tight-binding calculations \cite{schulzAtomisticAnalysisImpact2015,chaudhuriMultiscaleSimulationsElectronic2021}. This asymmetry in localization arises from the difference in effective masses between electrons and holes in the III-nitrides ($m_e^* \approx 0.2m_0$ and $m_h^* \approx 1.8m_0$). This finding is crucial for understanding recombination dynamics, as we discuss later. As a result of the weak impact of disorder on the electron wave function, our investigation centers on the effects of hole localization on carrier-recombination dynamics.

\begin{figure}[htp]
    \centering
    \includegraphics[width=0.5\textwidth]{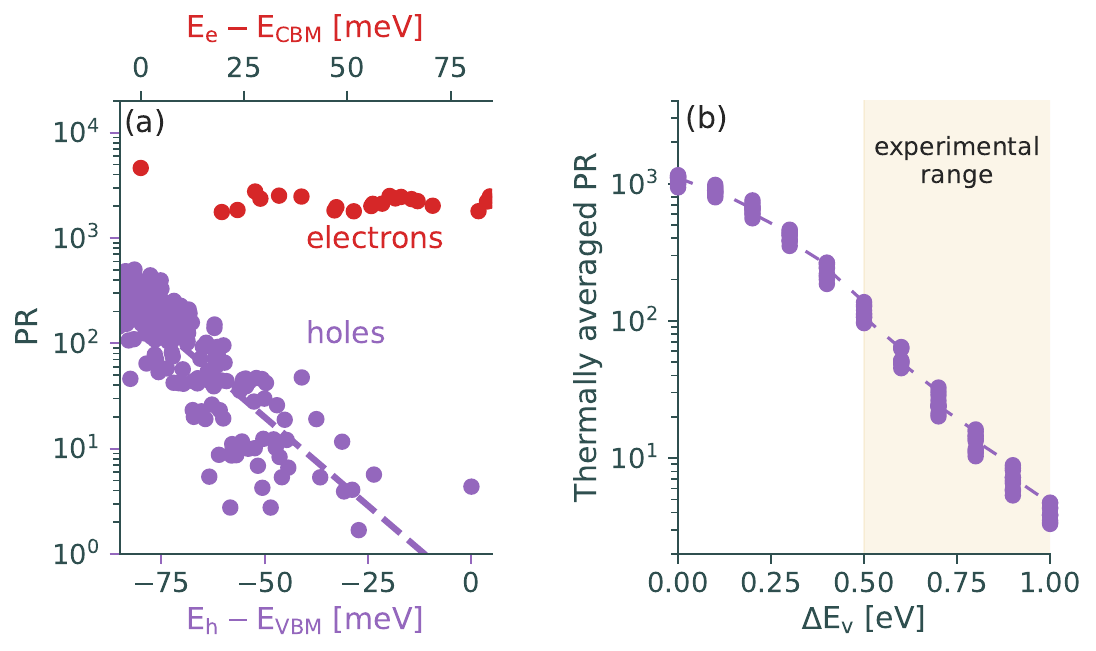}
    \caption{Impact of disorder on wave-function localization. (a) Participation ratio of electron and hole wave functions in an In$_{0.15}$Ga$_{0.85}$N alloy, calculated with first-principles material parameters. A smaller participation ratio indicates a more strongly localized wave function. (b) Larger hole-scattering potentials $\Delta E_v$ lead to more strongly localized holes in InGaN because of stronger compositional disorder.}
    \label{fig:2}
\end{figure}

To investigate the influence of hole localization on recombination, we controlled the disorder strength by tuning the \textcolor{black}{alloy-scattering potential. In our calculations, we vary the tune the hole scattering potential from 0.0 eV to 1.0 eV, while fixing the electron-scattering potential to the theoretical CB offset of 2.3 eV.} To quantify the degree of localization, we employ the thermally averaged participation ratio (PR), which measures the number of sites over which the wave functions are extended. We define the PR as $\left[\int d^3\textbf{r} \abs{\psi(\textbf{r})}^2\right]^2/\int d^3\textbf{r}  \abs{\psi(\textbf{r})}^4$. \textcolor{black}{According to this definition, a state localized to a single site has PR equal to the differential volume element, and a state fully extended over the simulation volume has PR equal to the simulation volume.} \textcolor{black}{Using scattering potentials equal to first-principles CB and VB offsets, we find that electrons are extended within the simulation cell but holes near the valence-band maximum are strongly localized, with their PR decreasing exponentially near the band edge (Figure \ref{fig:2}(a)). Moreover, increasing the hole-scattering potential leads to 
 stronger hole localization due to the heightened disorder strength (Figure \ref{fig:2}(b)).}

\subsection{Localization reduces recombination rates}

\begin{figure}[htp]
    \centering
    \includegraphics[width=0.5\textwidth]{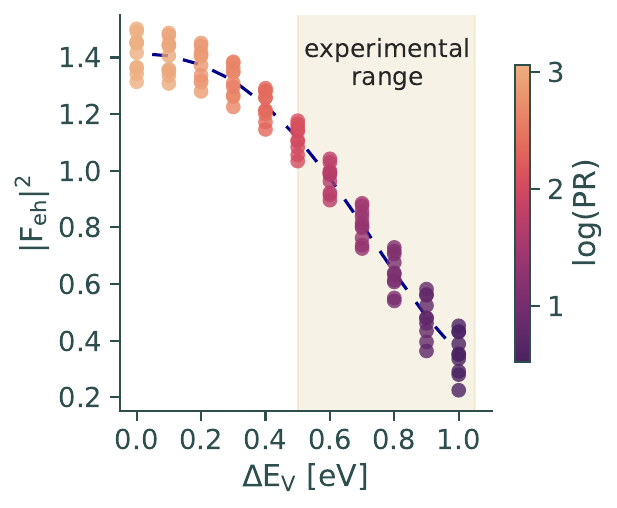}
    \caption{Impact of disorder on radiative recombination. Strong hole localization as a consequence of stronger disorder reduces the wave-function overlap, decreasing the rate of radiative recombination. The colors indicate the thermally averaged participation ratio of the hole wave functions; darker colors correspond to stronger localization.}
    \label{fig:3}
\end{figure}

Our findings demonstrate that compositional disorder in III-nitrides leads to  a decrease in the overlap of electrons and holes $\abs{F_{eh}}^2$, primarily driven by the asymmetry in their effective masses. Figure \ref{fig:3} illustrates the dependence of $\abs{F_{eh}}^2$ on hole localization in In$_{0.15}$Ga$_{0.85}$N alloys. \textcolor{black}{The reason $\abs{F_{eh}}^2 >1$ for $\Delta E_v = 0$ is that there are strain fluctuations in our simulation that reduce the symmetry of the alloy even for $\Delta E_v =0$ and enable optical transitions that are symmetry-forbidden in a virtual crystal.} As holes, on average, become more localized, their spectral weight is transferred away from the $\Gamma$ point, reducing their coupling with extended electrons whose spectral weight is highly concentrated near $\Gamma$. Figure S1 of the Supplemental Material (SM) shows that artificially localizing electrons alongside holes increases the wave-function overlap \cite{SM}. Consequently, the asymmetry in effective masses within the III-nitrides leads to a decrease in the rate of radiative recombination with stronger disorder.

\begin{figure*}[htp]
    \includegraphics[width=\textwidth]{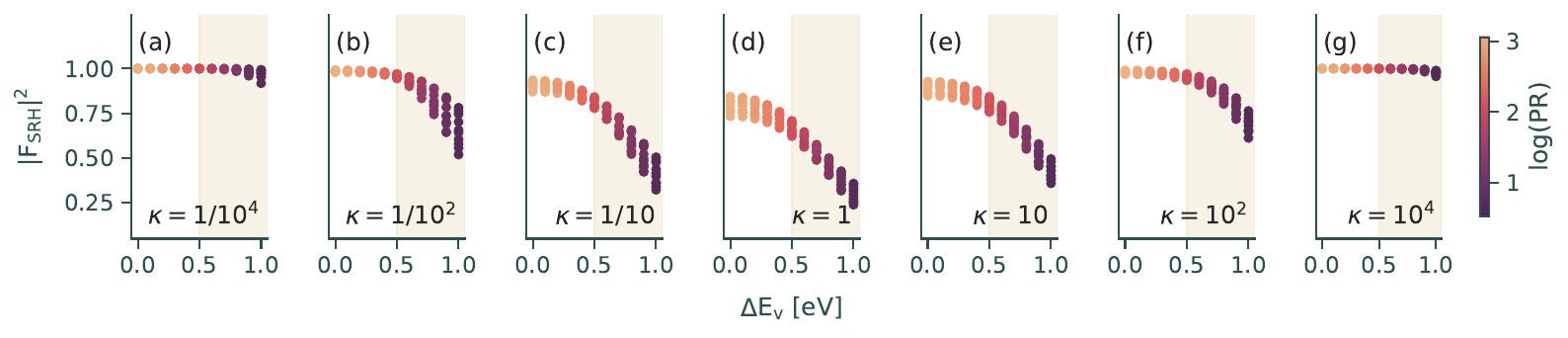}
    \caption{Impact of disorder on Shockley-Read-Hall recombination. Stronger hole localization as a consequence of stronger disorder reduces the rate of SRH recombination over defects with symmetric electron and hole capture coefficients ($\kappa \sim 1$). However, localization has little to no effect on recombination over defects with asymmetric capture coefficients ($\kappa \ll 1$; $\kappa \gg 1$). The colors indicate the thermally averaged participation ratio of the hole wave functions; darker colors correspond to stronger localization. The shaded region shows the range of experimentally measured values of the InN/GaN VB offset.}
    \label{fig:4}
\end{figure*}

On the other hand, the effect of disorder on non-radiative recombination depends on the $\kappa$ of the defect over which recombination occurs. Figure \ref{fig:4} shows the influence of carrier localization on $\abs{F_{SRH}}^2$ for recombination over defects with varying $\kappa$ values. For $\kappa$ close to unity, hole localization reduces the SRH integral by reducing the probability of finding an electron and hole at a defect site. In contrast, for extreme values of $\kappa$, hole localization has no effect on the SRH integral. This is because the SRH cycle is limited by multi-phonon emission rather than hole localization, as the presence of a hole at a defect site is always finite due to thermal occupation of extended states. Since defects with symmetric capture coefficients are typically the most efficient non-radiative recombination centers, overall, disorder tends to reduce the rate of SRH recombination.

\subsection{Power-law scaling of radiative and SRH recombination}
\label{sec:power-law}

\begin{figure*}[htp]
    \includegraphics[width=\textwidth]{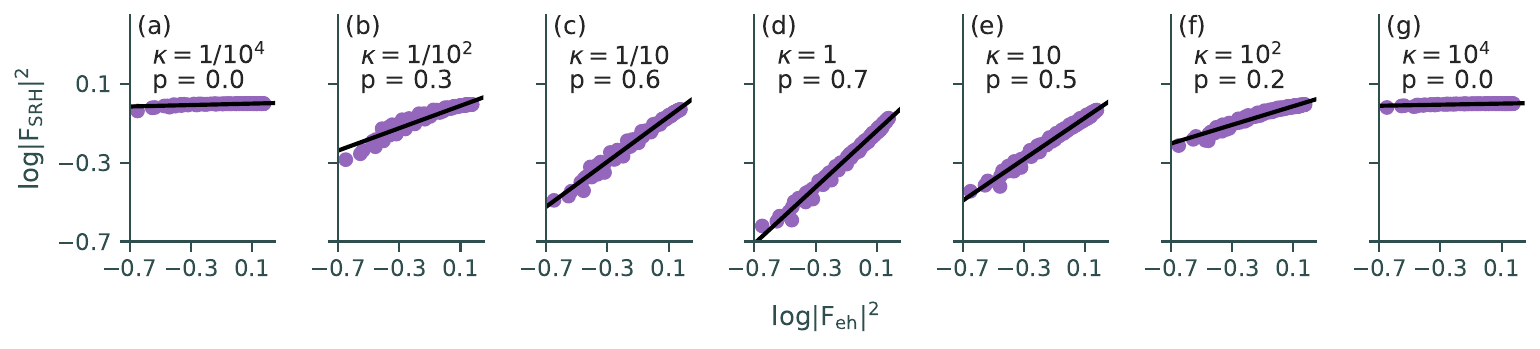}
    \caption{Scaling relation between radiative and SRH recombination. The integral for SRH recombination is related to the electron-hole wave-function overlap probability as a power law of the form $\abs{F_{SRH}}^2 \propto \abs{F_{eh}}^{2p}$. The scaling exponent $p$ (slope in log-log plot) depends on the $\kappa$ of the defect over which recombination occurs. Each panel corresponds to a different value of $\kappa$. The rate of SRH recombination and the rate of radiative recombination are strongly correlated for $\kappa$ close to one.}
    \label{fig:5}
\end{figure*}

Previous studies have assumed that carrier localization only reduces the $B$ coefficient but has no impact on the $A$ coefficient \cite{aufdermaurEfficiencyDropGreen2016, jonesImpactCarrierLocalization2017, tannerPolarGaGa2020}. We find that this assumption is not justified. Carrier localization reduces the rates of both radiative and SRH recombination in a correlated manner, and the correlation follows a power law of the form, 
\begin{equation}
    \abs{F_{SRH}}^2 = s(\kappa) \abs{F_{eh}}^{2p(\kappa) - 2}.
\end{equation}
In Figure \ref{fig:5}, we show this scaling relation and the power law exponent $p$ explicitly for various values of $\kappa$. We find the following approximate expressions for $s(\kappa)$ and $p(\kappa)$: $s(\kappa) \approx -0.67 \exp\left(-\frac{{(\log_{10}\kappa)^2}}{{4.0}}\right) + 1$ and $p(\kappa) \approx 0.71 \exp\left(-\frac{{(\log_{10}\kappa)^2}}{{4.6}}\right)$. If $\kappa$ is close to unity, the rate of non-radiative recombination is proportional to the probability of finding both an electron and hole at a defect site, therefore $|F_{\text{SRH}}|^2$ is proportional to $|F_{eh}|^2$. In contrast, $|F_{\text{SRH}}|^2$ is independent of $|F_{eh}|^2$ for extremely large or extremely small values of $\kappa$ because the rate of non-radiative recombination is limited by the carrier-capture process. In Figures S2 and S3 of the SM, we also show that assuming non-random defect distributions does not change our conclusions. The power-law relationship between radiative and non-radiative recombination, as demonstrated in our study with varying hole localization, seems to be a general feature of recombination in III-nitrides. Other authors have observed it for varying quantum-well thickness and composition, showing that polarization fields have a similar effect \cite{davidFieldassistedShockleyReadHallRecombinations2017,davidCompensationRadiativeAuger2019,davidReviewPhysicsRecombinations2019}.

\subsection{Analysis of point-defect tolerance}

The IQE ultimately depends on the ratio of the radiative to the total recombination rate. We capture this competition by considering the impact of wave-function overlap on the recombination rates,
\begin{linenomath}
\begin{align}
\text{IQE} &= \left(\frac{R_{SRH}}{R_{rad}} + 1\right)^{-1} = \left(\frac{{|F_{\text{SRH}}}|^2}{{|F_{eh}|}^2} \frac{{c_{tot}}N_Tn}{{B_0 n^2}} + 1\right)^{-1} \nonumber \\&= \left(s(\kappa) |F_{eh}|^{2p(\kappa) - 2} \frac{{c_{tot} N_T}}{{B_0 n}} + 1\right)^{-1}. \label{eq:IQE}
\end{align}
\end{linenomath}
Here, $c_{tot}$ is the total carrier-capture coefficient by point defects calculated without localization effects, \textit{e.g.}, at the level of density-functional theory, $N_T$ is the defect density, and $B_0$ is the bulk radiative recombination coefficient also calculated without localization effects. $|F_{\text{SRH}}|^2$ and $|F_{eh}|^2$ introduce corrections due to disorder or polarization fields for the case of polar quantum wells. Since we are interested in the current regime where defect-mediated SRH recombination dominates, we neglect third-order Auger-Meitner recombination (AMR) \cite{kioupakisIndirectAugerRecombination2011,matsakisRenamingProposalAuger2019}, however we later verify that the presence of AMR does not change our conclusions using experimentally measured coefficients. We have simplified the IQE expression by rewriting the SRH integral using the power-law relation discussed in Section \ref{sec:power-law}. 

We now \textcolor{black}{argue} that the carrier localization does not lead to point-defect tolerance in InGaN. In order for carrier localization to give rise to defect tolerance, the IQE, at a given carrier density and a given defect density, would have to increase with stronger localization. Instead, we find that stronger localization \textit{decreases} the IQE at a fixed carrier density (Figure \ref{fig:6}(a)). To evaluate the IQE expression in equation (\ref{eq:IQE}), we varied $|F_{eh}|^2$ from 0.1 to 1.5 and used a DFT-level recombination coefficient of $B_0 = 6 \times 10^{-11} \, \text{cm}^3 \, \text{s}^{-1}$. For the $A$ coefficient, we assumed a fixed defect density such that the product $c_{tot} N_T \equiv 10^7 \, \text{s}^{-1}$. The absence of defect tolerance can be understood by observing that the scaling exponent $p$, which governs the power-law relation between radiative and SRH recombination, is always positive. Defect tolerance would require either $p$ to be negative or if $p$ equals zero then $\abs{F_{eh}}^2$ would have to increase with stronger localization. \textcolor{black}{As neither condition is met, this analysis challenges that idea that InGaN is tolerant to point defects.}

\subsection{Why does the IQE improve with increasing indium content?}

The improvement of the IQE with increasing indium concentration (Figure \ref{fig:def-tol}) has thus far been attributed to the enhanced defect tolerance of In-containing alloys. \textcolor{black}{Can another mechanism explain this phenomenon?} One possibility is that point defects do not play a significant role in SRH recombination for the measured samples, and the observed increase in the IQE can be explained completely in terms of greater energetic screening of threading dislocation cores with increasing indium concentration. We believe this scenario to be highly unlikely since first-principles calculations have shown that point defects in the nitrides will form and contribute to SRH recombination even in samples of the highest quality \cite{alkauskas2016role, wickramaratne2020deep, lyons2021first}. There is also increasing evidence that the majority of non-radiative losses in the active region of InGaN LEDs occurs through point defects rather than threading dislocations \cite{armstrongDefectreductionMechanismImproving2015, hallerBuryingNonradiativeDefects2017, hallerGaNSurfaceSource2018, polyakovEffectsMeVElectron2020, pivaDefectIncorporationIncontaining2020, roccatoEffectsQuantumwellIndium2021}. 

\begin{figure}[htp]
    \centering
    \includegraphics[width=0.5\textwidth]{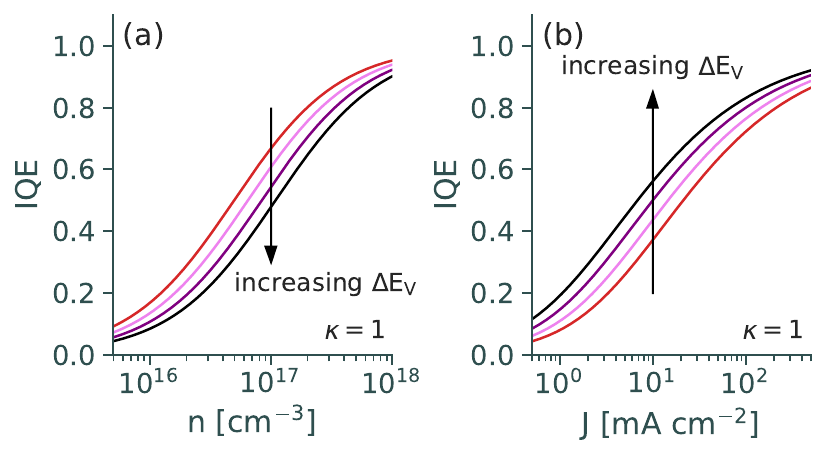}
    \caption{Disorder enhances IQE at a given current density. The IQE as a function of the carrier density $n$ versus current density $J$, using the scaling relation between $\abs{F_{SRH}}^2$ and $\abs{F_{eh}}^2$ that we calculated for $\kappa = 1$. (a) Stronger carrier localization due to stronger disorder decreases $\abs{F_{eh}}^2$ more quickly than $\abs{F_{SRH}}^2$, reducing the IQE at a given carrier density. (b) However, at a fixed current density, carrier localization increases the IQE by increasing the carrier density required to obtain a given current density, promoting bimolecular radiative recombination over mono-molecular SRH recombination, leading to an apparent increase in the IQE as the disorder becomes stronger.}
    \label{fig:6}
\end{figure}

Instead, the increase in the IQE with increasing indium concentration can be explained in terms of the fact that experiments are often performed under constant current conditions. Under constant current, higher indium-content emitters operate at higher carrier densities due to their slower recombination rates. An increase in the carrier density leads to the promotion of bimolecular radiative recombination ($R_{rad} \propto n^2$) over monomolecular SRH recombination ($R_{SRH} \propto n$), giving rise to an \textit{apparent} improvement of the IQE that can be mistaken for defect tolerance. We show this phenomenon explicitly in Figure \ref{fig:6}(b), where the IQE at a fixed current density increases with stronger localization, despite no change in the defect density. For this calculation, we assume $\kappa = 1$ since defects with symmetric capture coefficients tend to be the most active recombination centers but later we explicitly verify our conclusions with experimentally measured recombination coefficients, including the $C$ coefficient. (In Appendix \ref{app:radiativecapture}, we explain that our conclusions do not change in the limits $\kappa \ll 1$ or $\kappa \gg 1$. In these limits, multi-phonon emission is inefficient and band-to-band recombination competes with radiative capture by point defects, which scale linearly with each other.) Therefore, although carrier localization decreases the quantum efficiency at a given carrier density, it increases the quantum efficiency at a given current density, which is the relevant quantity for experimental measurements. As we discuss later in the text, this is the opposite of what occurs at higher current densities in the efficiency-droop regime.

\begin{figure}[htp]
    \centering
    \includegraphics[width=0.5\textwidth]{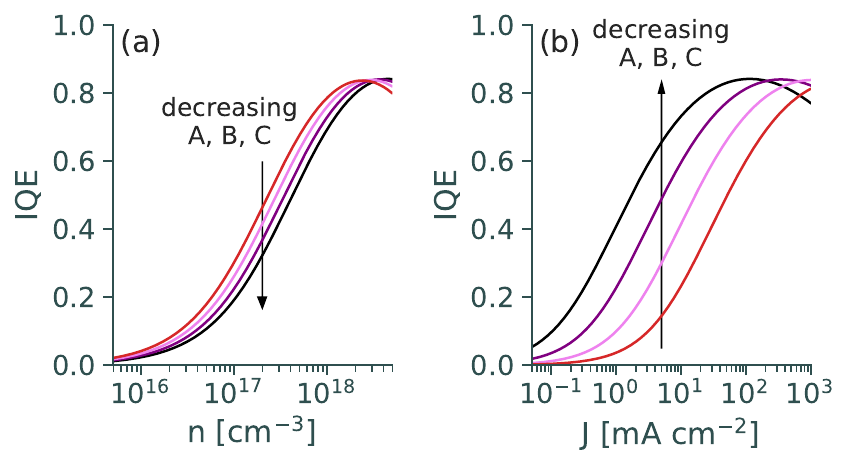}
    \caption{IQE enhancement in experimental systems that are representative of commercial LEDs. The IQE as a function of the carrier density $n$ versus current density $J$, modeled using empirical scaling relations between the $A$, $B$, and $C$ coefficients \cite{davidReviewPhysicsRecombinations2019}. (a) Carrier localization and polarization fields decrease $B$ more quickly than $A$, reducing the IQE at a given carrier density. (b) However, at a fixed current density, slower recombination dynamics increases the IQE by increasing the carrier density required to obtain a given current density, promoting radiative recombination over SRH recombination.}
    \label{fig:7}
\end{figure}

We now show that our proposed mechanism applies to commercial LEDs as well, despite the presence of additional factors such as carrier separation by polarization fields, Auger-Meitner recombination, variations in band gap and quantum-well thickness, and various types of defects contributing to non-radiative recombination. To account for these factors in our analysis, we use the empirical scaling relation between the $A$, $B$, and $C$ coefficients of samples of various thicknesses and compositions that are representative of commercial LEDs  \cite{davidReviewPhysicsRecombinations2019}. \textcolor{black}{We note that we do not explicitly model the screening of polarization fields because their impact on recombination coefficients in the carrier-density range of interest is rather small;  nevertheless, our use of scaling relations implicitly accounts for their effects on the competition between radiative and non-radiative recombination} \cite{davidReviewPhysicsRecombinations2019}. In Figure \ref{fig:7}, we present the IQE as a function of both carrier and current density, for $B$ coefficients ranging from $10^{-13} \, \text{cm}^3 \, \text{s}^{-1}$ to $10^{-11} \, \text{cm}^3 \, \text{s}^{-1}$. We observe that the IQE at a given carrier density decreases as the $B$ coefficient decreases, while the IQE at a given current density has the opposite behavior.

Finally, we revisit Figure \ref{fig:def-tol} and explicitly calculate the IQE to show that our mechanism can explain the improvement in the IQE with increasing indium concentration. In order to compare against experiments for which the total recombination rates are not known, we calculated the IQE for various current densities (1, 10, and 100 mA/cm$^2$). We used accurate $B$ coefficients for InGaN quantum wells that were calculated by diagonalizing atomistic tight-binding Hamiltonians parameterized to first-principles DFT calculations \cite{mcmahonAtomisticAnalysisAuger2022}. To obtain the $A$ and $C$ coefficients from the $B$ coefficient, we used empirical power-law relations measured for quantum-well designs that are representative of commercial LEDs \cite{davidReviewPhysicsRecombinations2019}. Our calculations, shown in Figure \ref{fig:def-tol}, clearly show that the improvement in the IQE with increasing indium concentration can be explained in terms of differences in carrier density rather than suppression of diffusion to non-radiative centers by carrier localization. The differences in carrier density can occur because of stronger carrier localization or stronger polarization fields, the latter of which has a greater effect as it can influence the recombination rates by orders of magnitude. We believe that further experimental work, using differential-lifetime-based techniques that give simultaneous access to the IQE and the operating carrier density \cite{davidReviewPhysicsRecombinations2019, liCarrierDynamicsBlue2023}, is warranted to verify the mechanism that we have proposed. 

\subsection{Consequences to devices}

While reducing the overlap of carrier wave functions can enhance the IQE at low current densities, it has drawbacks for high-power device performance. At higher current densities, decreased overlap favors third-order Auger-Meitner recombination over bimolecular radiative recombination \cite{kioupakisInterplayPolarizationFields2012,liCarrierDynamicsBlue2023}, shifts the maximum efficiency to lower current densities, and worsens the problematic hue shift of LEDs \cite{pantOriginInjectiondependentEmission2022}. This finding cautions against relying solely on comparisons at a single carrier-injection level if evaluating the efficiency of different devices, especially with regards to notions of defect tolerance. Moreover, it is crucial to recognize that reducing the carrier overlap overall decreases the IQE at a given carrier density, therefore the IQE is in fact highly sensitive to the defect density. This is consistent with observations that reducing the point-defect density dramatically improves the maximum IQE of compositionally disordered LEDs \cite{armstrongDefectreductionMechanismImproving2015, hallerBuryingNonradiativeDefects2017, hallerGaNSurfaceSource2018, polyakovEffectsMeVElectron2020, pivaDefectIncorporationIncontaining2020, roccatoEffectsQuantumwellIndium2021}. Nevertheless, there may be applications, such as low-power microLEDs with large surface recombination velocities \cite{sheenHighlyEfficientBlue2022}, where reducing the wave-function overlap, e.g., by promoting carrier localization or increasing the thickness of polar quantum wells, could be advantageous for improving the low-current quantum efficiency.

\section{Conclusion}

To conclude, we have addressed the longstanding controversy of whether compositional disorder in InGaN leads to defect tolerance. We do not find evidence that carrier localization is responsible for the highly efficient nature of InGaN, as commonly believed. A reduction in wave-function overlaps and recombination rates accompanied by an increase in the carrier density, rather than suppression of diffusion, can explain the enhancement of the light-emission efficiency with increasing indium concentration. \textcolor{black}{We believe there is good motivation for further experimental work using advanced techniques that give simultaneous access to the carrier density and the recombination rates in the active region to verify this mechanism.} At higher powers, the same mechanism exacerbates efficiency droop and worsens the hue shift of LEDs. Naturally, the maximum quantum efficiency depends strongly on the overall defect density regardless of the wave-function overlap, emphasizing the need to reduce defect concentrations. This work provides a theoretical framework applicable to other semiconductors for understanding the impact of disorder on recombination dynamics, paving the way for accurately assessing the impact of disorder on the performance of optoelectronic devices. 

\section*{Acknowledgements}
We thank Rob Armitage, Daniel Feezell, Mark Holmes, and Siddharth Rajan for useful discussions, and Kyle Bushick, Amanda Wang, and Xiao Zhang for their help editing this manuscript. This project was funded by the U.S. Department of Energy, Office of Energy Efficiency and Renewable Energy, under Award No. DE-EE0009163. Computational resources were provided by the National Energy Research Scientific Computing Center, a Department of Energy Office of Science User Facility, supported under Contract No. DEAC0205CH11231. N.P. acknowledges the support of the Natural Sciences \& Engineering Research Council of Canada Postgraduate Scholarship.

\appendix
\section{Derivation of $\abs{F_{SRH}}^2$}
\label{app:derivation}
We can write the rate of Shockley-Read-Hall (SRH) recombination (otherwise known as non-radiative recombination by multi-phonon emission) as being proportional to a quantity that we denote $\abs{F_{SRH}}^2$. This term is not strictly an overlap term but it involves overlap-like integrals of the wave functions, therefore we interchangeably refer to it as the SRH overlap term. To derive this term, we start from the generalized SRH recombination rate that accounts for non-uniformities in the charge density,

\begin{equation}
    R_{SRH} = n_T \int d^3 \textbf{r} \frac{c_n n(\textbf{r}) \times c_p p(\textbf{r})}{c_n n(\textbf{r}) + c_p p(\textbf{r})},
\end{equation}

where $c_n$ and $c_p$ are any given defect’s electron and hole capture coefficients, $n(\textbf{r})$ and $p(\textbf{r})$ are the spatially varying electron and hole carrier densities, and $n_T$ is the trap density. For this expression, we have assumed that traps are uniformly distributed, however we also explicitly verify this assumption by relaxing this assumption and allowing the trap density to vary spatially (see Figure S1 and S2 of the supplementary material). 

\textcolor{black}{The process of SRH recombination involves the capture of one carrier by a defect, followed by a subsequent capture of a second carrier. The atomistic defect-level physics of this phenomenon, including the multi-phonon-emission physics, is captured in the electron ($n$) and hole ($p$) capture coefficients, $c_{n,p}$. The capture coefficient is defined as, \cite{alkauskasFirstprinciplesTheoryNonradiative2014}
\begin{align}
    c_{n,p} = &V \frac{2\pi}{\hbar} g \abs{\bra{\phi_f}H_{ep}\ket{\phi_i}}^2 \sum_m w_m \sum_n \abs{\bra{\chi_f}\Delta Q\ket{\chi_i}}^2 \nonumber \\
    & \times \delta(\Delta E + m\hbar \Omega_i - n\hbar \Omega_f),
\end{align}
where $V$ is the simulation volume, $g$ is a degeneracy factor, $\phi_i$ is the initial atomistic wave-function prior to carrier capture and $\phi_f$ is the wave-function captured by the defect, $H_{ep}$ is the electron-phonon-coupling Hamiltonian, $m$ and $n$ index the vibrational modes of the excited electronic defect state, $w_m$ is a bosonic occupation factor, $\chi_i$ and $\chi_f$ are initial and final vibrational wave-functions of the defect, $\Delta Q$ is the change in normal coordinates of the defect before and after multi-phonon emission, and the $\delta$ function conserves energy during the electronic ($\Delta E$) and vibronic $(m\hbar \Omega_i - n\hbar\Omega_f)$ transitions. In principle, $c_n$ and $c_p$ should also vary spatially as the capture rate can depend on the local alloy configuration and nearest-neighbour bonding environment. However, explicit \textit{ab initio} calculations have shown that these effects average out once one considers that defects are found in a multitude of local environments, and one can accurately define \textit{effective} capture coefficients for the entire crystal \cite{wickramaratne2020deep}. As we will later show, if one is only interested in the \textit{correction} to the SRH rate relative to the virtual-crystal case, one only needs to know the ratio $\kappa \equiv c_p/c_n$ rather than the exact values of $c_n$ and $c_p$. Hence, in our formalism, the effects of the atomistic carrier-capture and multi-phonon-emission processes are embedded in $\kappa$. Since $\kappa$ can vary for different defects, we treat it as an input parameter of our theory, and vary it across a range of reasonable values from very large ($10^4$) to very small ($10^{-4}$). We then use our intuition that defect states with symmetric capture coefficients $\kappa \sim 1$ are most active for SRH recombination to extract conclusions for devices.}

\textcolor{black}{While the carrier capture coefficients describe the physics of multi-phonon-emission and defect-carrier capture, one also needs to consider the local charge densities $n(\textbf{r})$ and $p(\textbf{r})$ to evaluate the rate of SRH recombination. We construct the local charge density from the envelope functions calculated for the random alloy without defects in the supercell. The underlying assumption is that the defect states are so dilute that they do not significantly modify the charge density. Indeed, we have explicitly checked that modeling a single point defect with a localized deformation potential of order 1 eV has no discernible influence on the envelope functions. This assumption has also been used by other authors to model the SRH process in nitride quantum wells \cite{davidFieldassistedShockleyReadHallRecombinations2017}. It is important to note that this assumption may not hold for the case of charged defects, whose electrostatic field will cause a local redistribution of the charge density. In the defect literature, this correction goes under the name of the Sommerfeld enhancement-factor correction, and it should either be applied to the capture coefficient or the local charge density but not both. Since it is conventional to directly apply this correction factor in the calculation of the \textit{ab initio} capture coefficient \cite{alkauskasFirstprinciplesTheoryNonradiative2014}, we have excluded its effects in our calculation of $n(\textbf{r})$ and $p(\textbf{r})$ to avoid double counting. Other researchers employing similar methods as us have noted that charging effects are not essential for capturing the correct qualitative physics of nitride emitters \cite{davidFieldassistedShockleyReadHallRecombinations2017}.}  

\textcolor{black}{At this point, we perform further algebraic manipulations to rewrite $R_{SRH}$.} Under symmetric injection of electrons and holes, i.e., $\int d^3\textbf{r}\  n(\textbf{r}) = \int d^3\textbf{r}\ p(\textbf{r}) = N$, where $N$ is the total number of carriers, we can rewrite the electron and hole densities as $n(\textbf{r}) = N \delta n(\textbf{r})$ and $p(\textbf{r}) = N \delta p(\textbf{r})$, where the spatially varying parts are given by $\delta n(\textbf{r}) = \sum_{c} \abs{\psi_c(\textbf{r})}^2 f_c/\sum_c f_c$ and $\delta p(\textbf{r}) = \sum_v \abs{\psi_v(\textbf{r})}^2 (1-f_v) / \sum_v (1-f_v)$, where $c$ and $v$ index conduction and valence bands respectively, and $f$ is the Fermi-Dirac occupation probability. \textcolor{black}{Here, we stress that the summation over $c$ and $v$ occur over hundreds of eigenstates, allowing us to capture the effects of both localized and extended states.} Factoring out $c_n$, we can rewrite the SRH rate as,
\begin{equation}
    R_{SRH} = n_T c_n \int d^3\textbf{r} \frac{\delta n(\textbf{r}) \times \delta p(\textbf{r})}{\delta n(\textbf{r}) + (c_p/c_n) \delta p(\textbf{r}) }.
\end{equation}
For a system with translational invariance, the macroscopic charge density is spatially uniform and $\delta n(r) = \delta p(r) = 1/V$, thus $R_{SRH}^{VCA} = n_T(N/V)c_n/(1+c_p/c_n)$. We define $\abs{F_{SRH}}^2$ as the correction factor to the SRH recombination rate due to carrier localization effects, thus,
\begin{equation}
    \abs{F_{SRH}}^2 = \frac{R_{SRH}}{R_{SRH}^{VCA}} = V(1+\kappa) \int d^3 \textbf{r} \frac{\delta n(\textbf{r}) \times \delta p(\textbf{r})}{\delta n(\textbf{r}) + \kappa \delta p(\textbf{r})},
\end{equation}
where we have defined $\kappa \equiv c_p/c_n$. This is the expression that we have provided in the main text. \textcolor{black}{We reiterate that the effects of the atomistic carrier capture process, included details of the multi-phonon-emission physics, defect-charging effects, and details of the atomistic defect wave functions, are embedded in the $\kappa$ parameter.}\\

\section{Impact of radiative capture by defects on the IQE}
\label{app:radiativecapture}
In the main text, we showed that $\abs{F_{eh}}^2$ and $\abs{F_{SRH}}^2$ are almost linearly correlated for $\kappa \sim 1$, and completely uncorrelated for $\kappa \gg 1$ and $\kappa \ll 1$. We also showed that near-linear correlation between $\abs{F_{eh}}^2$ and $\abs{F_{SRH}}^2$ leads to an apparent defect tolerance at low current densities. In the extreme limits of $\kappa$, multi-phonon emission becomes prohibitively slow, in which case the dominant mono-molecular recombination process that competes with band-to-band radiative recombination is radiative capture by point defects, such as the $\text{C}_\text{N}$ impurity in GaN. This modifies the ABC model as,
\begin{equation}
    IQE = \frac{Bn^2}{An + B'n + Bn^2 + Cn^3},
\end{equation}
where $B'$ is the radiative-capture recombination coefficient. Similar to the $B$ coefficient, the $B'$ coefficient is also proportional to the electron-hole overlap $\abs{F_{eh}}^2$ as it is proportional to a momentum matrix element $\abs{p_{if}}^2$ governing defect-to-band or band-to-defect transitions. Therefore, our conclusions do not change even if band-to-band recombination competes with radiative capture by point defects rather than multi-phonon emission.